\documentclass{sig-alternate}

\usepackage{cite}

\graphicspath{{figures/}}
\DeclareGraphicsExtensions{.pdf}

\usepackage{url}

\usepackage{algpseudocode}

\newtheorem{defn}{Definition}

\usepackage{xspace}

\usepackage{ifthen}

\usepackage{listings}

\usepackage{booktabs}

\usepackage{numprint}
\npthousandsep{,}        

\mathchardef\mhyphen="2D 

\newcommand{\seqOf}[1]{\left[\,#1\,\right]}

\newcommand{\produces}{\mbox{$\rightarrow\;$}}
\newcommand{\synOr}{\mbox{$\;\,|\;\;$}}

\newcommand{\etal}{\emph{et al.}\@\xspace}

\newcommand{\msgShapes}{\mbox{$\varsigma$}}

\newcommand{\msgs}{\mbox{$\mathcal{M}$}}
\newcommand{\values}{\mbox{$\mathcal{V}$}}

\newcommand{\aMsgShape}{\mbox{$\sigma$}}

\newcommand{\inSymb}{\mbox{$?$}}

\newcommand{\outSymb}{\mbox{$!$}}

\newcommand{\protocols}{\mbox{$\mathcal{P}$}}

\newcommand{\choice}[2]{#1 + #2}
\newcommand{\contrSymb}{{\triangleright}}
\newcommand{\contractInteraction}[2]{#1 \,\contrSymb\, #2}
\newcommand{\declProt}[2]{\mbox{$#1\!=\!#2$}}

\newcommand{\decl}{D}
\newcommand{\interact}{I}
\newcommand{\specs}{S}

\newcommand{\var}{v}
\newcommand{\extend}[2]{{#1  \sim  #2}}
\newcommand{\inaction}{\mbox{\bf 0}}
\newcommand{\product}[2]{#1 \, * \, #2}

\newcommand{\vars}{V}
\newcommand{\standardInteraction}[2]{#1 . #2}

\newcommand{\origin}{o}
\newcommand{\protocolMdlAlphElement}[2]{#1 #2}

\newcommand{\ldapAddRes}{\msgShapeLDAP{AddRes}}
\newcommand{\ldapAddRq}{\msgShapeLDAP{AddRq}}
\newcommand{\ldapBaseVar}{\protoVariable{Base}}
\newcommand{\ldapBindRes}{\msgShapeLDAP{BindRes}}
\newcommand{\ldapBindRq}{\msgShapeLDAP{BindRq}}
\newcommand{\ldapDelRes}{\msgShapeLDAP{DelRes}}
\newcommand{\ldapDelRq}{\msgShapeLDAP{DelRq}}
\newcommand{\ldapModRes}{\msgShapeLDAP{ModRes}}
\newcommand{\ldapModRq}{\msgShapeLDAP{ModRq}}
\newcommand{\ldapSearchDone}{\msgShapeLDAP{SearchDone}}
\newcommand{\ldapSearchEntry}{\msgShapeLDAP{SearchEntry}}
\newcommand{\ldapSearchRq}{\msgShapeLDAP{SearchRq}}
\newcommand{\ldapSearchVar}{\protoVariable{Search}}
\newcommand{\ldapUnbindRq}{\msgShapeLDAP{UnbindRq}}
\newcommand{\msgShapeLDAP}[1]{\mbox{$\mathsf{#1}$}}
\newcommand{\protoVariable}[1]{\underline{#1}}

\newcommand{\dataStores}{\mbox{$\mathcal{D}$}}

\newcommand{\handleRqN}{\mbox{$\mathsf{handle \mhyphen rq}$}}

\newcommand{\Kaluta}{{\sc Kaluta}}

\newcommand{\IM}{{IdentityManager\xspace}}
\newcommand{\IMShort}{{IM\xspace}}

\begin{document}




\title{Enterprise Software Service Emulation: Constructing Large-Scale Testbeds}

\numberofauthors{4} 
%
\author{
%
%
\alignauthor Cameron Hine\\
       \affaddr{Swinburne University of Technology}\\
       \affaddr{Hawthorn, Victoria, Australia}\\
       \email{chine@swin.edu.au}
\alignauthor Jean-Guy Schneider\\
       \affaddr{Swinburne University of Technology}\\
       \affaddr{Hawthorn, Victoria, Australia}\\
       \email{jschneider@swin.edu.au}
\alignauthor Jun Han\\
       \affaddr{Swinburne University of Technology}\\
       \affaddr{Hawthorn, Victoria, Australia}\\
       \email{jhan@swin.edu.au}
\and
\alignauthor Steve Versteeg\\
       \affaddr{CA Research}\\
       \affaddr{CA Technologies}\\
       \affaddr{Melbourne, Australia}\\
       \email{steve.versteeg@ca.com}
}

\maketitle




\begin{abstract}
  Constructing testbeds for systems which are interconnected with large
  networks of other software services is a challenging task. It is
  particularly difficult to create testbeds facilitating evaluation of the
  non-functional qualities of a system, such as scalability, that can be
  expected in production deployments.
  %
  %
  \emph{Software service emulation} is an approach for creating
  such testbeds where service behaviour is defined by emulate-able
  models executed in an emulation runtime environment.
  We present (i) a meta-modelling framework supporting emulate-able
  service modelling (including messages, protocol, behaviour and
  states), and (ii) {\Kaluta}, an emulation environment able to
  concurrently execute large numbers (thousands) of service models,
  providing a testbed which mimics the behaviour and characteristics
  of large networks of interconnected software services.
  Experiments show that {\Kaluta} can emulate \numprint{10 000}
  servers using a single physical machine, and is a practical testbed
  for scalability testing of a real, enterprise-grade identity
  management suite. The insights gained into the tested enterprise
  system were used to enhance its design.
\end{abstract}

\section{Introduction}
\label{sec:intro}

Testing software systems which rely on interactions with large networks of
interconnected software services is difficult. Testbeds mimicking the
behaviour and characteristics of such environments are necessary for a {\em
  system under test} (SUT) to be evaluated in production-like conditions. The
scale of enterprise environments is quite difficult for to replicate in
testbeds as it is common for there to be tens-of-thousands of systems
operating within a single environment. Constructing testbeds representing such
scales is inadequately supported by existing techniques, yet is crucial for
evaluating non-functional attributes of systems, such as scalability, intended
to operate in such environments.



\begin{figure}[t]
  \centering
  \includegraphics[width=0.75\columnwidth]{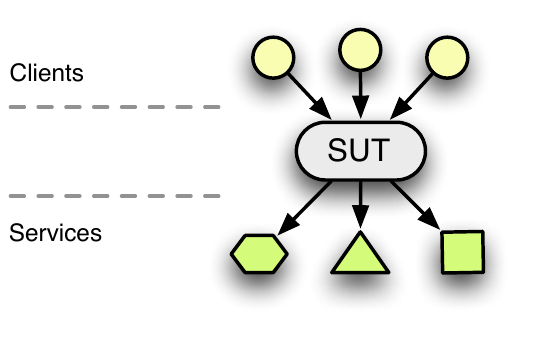}
  \caption{Polar Perspectives of Software Testbeds.}
  \label{fig:persp-test-env}
\end{figure}

A system under test's deployment environment can be considered from two polar
perspectives (cf. Fig.~\ref{fig:persp-test-env}): (i) the \emph{service}, or
service \emph{provider} perspective, and the (ii) \emph{client}, or service
\emph{consumer} perspective. From the former perspective, systems make
requests of an SUT whereas in the latter perspective, an SUT acts as the
client issuing requests to services offered by other systems in the
environment.


Testing systems from a service \emph{provider} perspective (often referred to
as {\em load testing}) is well supported by existing tools. For example, load
generating performance testing tools (e.g., SLAMD, HP Load Runner, and Apache
JMeter) can generate scalable amounts of client load, imitating large numbers
of concurrent users, allowing evaluation of a SUT's performance and
scalability under those loads. However, these tools can, in general, not
operate in ``reverse'' mode, that is, {\em responding} to incoming requests
from a SUT.


In order to serve requests from clients, a service may need to make requests
to other third party services, thereby being a service \emph{consumer} as
well. Testing a SUT from a {\em consumer} perspective requires an environment
containing entities imitating the behaviour of software services the SUT makes
requests to. The mock objects approach~\cite{Freeman2004} allows developers to
define superficial implementations of external services.
%
%
However, by providing service imitations hooked directly into a SUT's code,
calls to the host environment, such as file and networking services, are
typically bypassed and, consequently, are inappropriate for system testing
purposes.


Another approach, widely adopted by industry, is the use of system-level
virtual machines (VMs),
to create testing environments. VM environments are well suited for system
testing as their behaviour is essentially equivalent to their real
counterparts. Scalability, however, is a limiting factor. While it varies
depending on workloads, the general rule of thumb is a virtual CPU to physical
core ratio in the order of ten to one is the practical upper
limit~\cite{Sanchez:11}. This means that clusters of high-end machines are
required to host environments containing tens-of-thousands of VMs.


Contrasting load testing that employs an SUT {provider} perspective,
\emph{service emulation} is an approach we propose for constructing
large-scale testbeds that imitate interaction behaviour of software services,
thereby catering to the needs of a service \emph{consuming} SUT. There are two
key elements in the service emulation approach: (i) modelling {\em
  approximations} of real service behaviours to an adequate fidelity, that is,
to a level accurate enough that the desired test scenarios can be carried out,
and (ii) an emulation environment capable of simultaneously executing many
such models, presenting an appearance and behaviour similar to a real
enterprise service environment. The idea being that run-time properties of an
SUT, such as scalability and performance, can be evaluated whilst interacting
with emulated, rather than real, services.

In this paper, we present three primary contributions resulting from our work
on service emulation. In Section~\ref{sec:approach}, we present a (i) layered
{\em service meta-model} enabling service modelling. Flexibility is a key
feature of this model as it is paramount so that the diverse needs of
different testing scenarios can be accommodated. (ii) {\Kaluta}, an emulation
environment supporting scalable service model execution and run-time
interaction with real service consuming systems, that is SUTs, is also
presented in Section~\ref{sec:approach}. (iii) An empirical evaluation follows
in Section~\ref{sec:evaluation}, investigating {\Kaluta}'s scalability and
resource consumption (RQ1), comparing it with current industry best practises
(VM based testbeds) (RQ2), and its practicality -- applying it to a real
industry testing scenario (RQ3). Also included in this paper is a concrete
industry scenario framing our work in a practical context
(Section~\ref{sec:ind-scen}). Related work is discussed in
Section~\ref{sec:related-work} while Section~\ref{sec:conclusion} concludes
the paper, summarising the major results and suggesting areas for future work.

\section{Industry Scenario}
\label{sec:ind-scen}


CA {\IM} ({\IMShort}){\footnote{\small
    \url{http://www.ca.com/us/products/ca-identity-manager.html}}} is an
enterprise-grade identity management suite supporting management and
provisioning of users, identities and roles in large organisations covering a
spectrum of different endpoint types. It is typically deployed into large
corporations, such as banks and telecommunications providers, who use it to
manage the digital identities of personnel as well as to control access of
their vast computational resources and services. {\IMShort} invokes services
on the endpoints it manages to perform tasks such as:
\begin{itemize}
\item \textbf{Endpoint Acquisition:} IM obtains and validates \mbox{login}
      credentials to administer an endpoint.
\item \textbf{Endpoint Exploration:} Retrieve an endpoint's identity
  objects.
\item\textbf{Add Account:} Add user credentials to an endpoint
  permitting access.
\item\textbf{Modify Account:} Change some fields of an account.
\end{itemize}


{\IMShort} is regularly deployed into environments requiring the management of
tens-of-thousands of systems. Evaluating {\IMShort}'s run-time properties is
important so that it can be confidently deployed into large production
environments. Performing these evaluations with every release can detect
scalability issues introduced by new features or other code modifications.
Furthermore, {\IMShort} deployment teams can use results to guide improvements
in production environments. However, as results obtained in small environments
do not necessarily translate to production scale reality, evaluations require
a testbed containing tens-of-thousands of endpoints. The behaviour of the
testbed must be accurate enough so that {\IMShort} can carry out its key
operations (as listed above). Finally, in order to detect run-time issues
related to networking and other operating system services, the testbed cannot
rely on hooks into {\IMShort}'s code to act as its environment.





\section{Approach}
\label{sec:approach}


Service emulation~\cite{Hine2009} is an approach to constructing large-scale
testbeds which mimics interaction behaviour and characteristics of real
production environments. Ideally, QA teams can use service emulation to
uncover issues which otherwise remain hidden until triggered in production.

\begin{figure}[tb]
  \centering 
  \includegraphics[width=0.75\columnwidth]{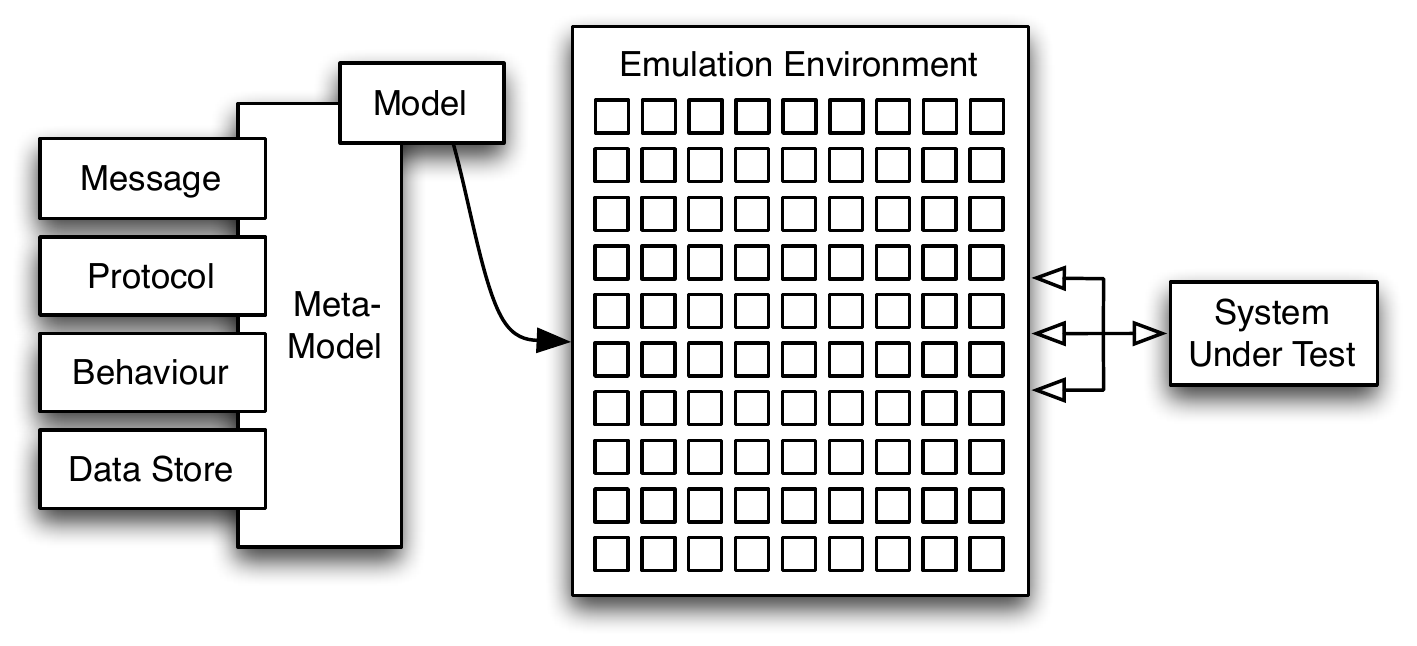}
  \caption{Service Emulation Approach.}
  \label{fig:service-emulation}
  \vspace{-0.4cm}
\end{figure}

The service emulation approach is based on the idea that \emph{approximations}
of service behaviour can be described using light-weight models which are
executed in an emulation environment to provide run-time behaviour.
Fig.~\ref{fig:service-emulation} outlines the key elements of our approach: a
service \emph{meta-model}, emulate-able service \emph{models} defined in terms
of the meta-model, and an \emph{emulation environment} which executes many
service models simultaneously and also handles communication between emulated
service models and SUTs.



\subsection{Service Meta-Model and Models}
\label{sec:meta-models}


Our service meta-model is a modelling framework capable of expressing the
approximate behaviour of real software services (e.g., LDAP). The service
meta-model is organised into four layers: the message model defines
\emph{what} is communicated between systems; the protocol model specifies
\emph{when} certain messages may be sent; the behaviour model defines
\emph{how} services respond to requests; and the data store persists updates
to the service state.



\subsubsection{Messages}


Remote software systems interact with one another by exchanging messages over
a computer network. During transmission, messages can be encoded in a variety
of different ways: 
encapsulated within HTTP chunks or following the ASN.1 basic encoding rules
(BER) specification. Our meta-model introduces a standard message format for
service models. Various encodings are supported by translating between these
encodings and the standard format.

Messages are defined (cf. Definition~\ref{def:msg-meta-mdl}) as having two
parts: a shape (or type), and a value. Message values may have different
structures, which we model as \emph{value} sequences. Within value sequence we
allow \emph{associated} values, associating a label with a value, to help
identify values in sequences. Service specific treatment of values is
supported through a generic \emph{base} value option.

\begin{defn}[Message Meta-Model]
  \begin{align*}
    \msgs &= \msgShapes \times \seqOf{\values}
  \end{align*}
  where {\msgs} denotes the set of messages, {\msgShapes} the set of message
  shapes, and {$\seqOf{\values}$} the set of value sequences, respectively.
  \label{def:msg-meta-mdl}
\end{defn}

\subsubsection{Protocols}


Protocols govern the rules for communication between software systems. For the
purpose of our work, we consider a protocol to define the \emph{temporal
  order} of messages which are valid for exchange throughout interactions.
Protocol models can be used to guide service model behaviour and can ensure
that emulated service transmissions are valid throughout interactions with
systems under test.




The abstract syntax for our protocol meta-model is presented in
Definition~\ref{def:abs-syn-prot-mdl}. For a detailed description of each
element, the reader is referred to \cite{Hine2010}.




\begin{defn}[Abstract Syntax]
  \begin{small}
    \begin{align*}
      \specs &\produces \decl &\textit{Specification} &&\decl
      &\produces \declProt{\var}{\protocols} \textbf{ and } \decl &
      \textit{Multi Declaration} \\
      &\synOr \protocols &&&&\synOr
      \declProt{\var}{\protocols}\textbf{
        in } \protocols &\textit{Single Declaration} \\[0.2cm]
    \end{align*}
    \vspace{-12mm}
    \begin{align*}
      \protocols &\produces \product{\protocols}{\protocols}
      &\textit{Product} &&\interact &\produces
      \choice{\interact}{\interact} &\textit{Choice} \\
      &\synOr \extend{\protocols}{\protocols} &\textit{Extension}
      &&&\synOr \standardInteraction{\origin\aMsgShape}{\protocols}
      &\textit{Standard Interaction}\\
      &\synOr \interact &\textit{Interaction} &&&\synOr
      \contractInteraction{\origin\aMsgShape}{\protocols}
      &\textit{Contractive Interaction}\\
      &\synOr  \var  &\textit{Variable}\\
      &\synOr  \inaction  &\textit{Inaction}
  \end{align*}
\end{small}
\vspace{+0.05cm} \noindent
where $\var$ is an element of $\vars$, the set of protocol variables;
$\origin$ denotes the direction of a message (transmission or reception); and
$\aMsgShape$ is an element of the set of message shapes $\msgShapes$.
  \label{def:abs-syn-prot-mdl}
\end{defn}



\vspace{-0.5cm}
Message \emph{interactions} are the fundamental events 
for protocols. An interaction is either the reception ({\inSymb}) or
transmission ({\outSymb}) of a message from the perspective of the service
being modelled. Our protocol meta-model defines message interactions using
message shapes rather than message content. This allows protocol models to
abstract away from message content details and focus on temporal
concerns. Therefore, the protocol meta-model's interaction events are defined
as direction/message shape pairs.
%
%
Interaction events are annotated with \emph{continuations}, defining what
interactions, if any, are valid \emph{after} the corresponding interaction
event. Furthermore, the protocol meta-model enables non-determinism by
incorporating \emph{choice} between interaction events.

The protocol meta-model includes operations for standard (\emph{product}) and
subservient (\emph{extension}) parallelism. In product compositions,
interaction sequences of protocols are treated independently. In contrast,
extension(s) of a given protocol are not fully independent and may be
terminated by interaction events occurring in the parent protocol, denoted as
protocol \emph{contraction}. These operations together enable concise
specifications of subservient parallelism exhibited in service protocols such
as LDAP. 
Protocol \emph{specifications}, \emph{declarations} and \emph{variables}
facilitate definition of more complex protocols~\cite{Hine2010}.






Figure~\ref{fig:ldap-prot} illustrates the service protocol meta-model by
modelling the LDAP directory server protocol. To enhance readability, all
occurrences of protocol variables are underlined. We specify the base
protocol functionality in $\ldapBaseVar$, the functionality of searching in
$\ldapSearchVar$, and extend $\ldapBaseVar$ with $\ldapSearchVar$ whenever a
new search request is received.
%
%
In order to enable the non-blocking of an LDAP server, in the context of
processing administrative and data modifying requests, the $\ldapBaseVar$
protocol is extended with protocol specifications encoding the appropriate
responses. Contractive interactions are used to terminate any pending
operations when a {\small $\ldapBindRq$} or {\small $\ldapUnbindRq$} request
is received.


\begin{figure}[tb]
 {\small
  \begin{align*}
    \ldapBaseVar = & \; \contractInteraction
    {\protocolMdlAlphElement{?}{\ldapUnbindRq}} {\inaction} \;+\;
    \contractInteraction{\protocolMdlAlphElement{?}{\ldapBindRq}}
    {\standardInteraction {\protocolMdlAlphElement{!}{\ldapBindRes}}
      {\ldapBaseVar}} \\ & \;+\;
    \standardInteraction{\protocolMdlAlphElement{?}{\ldapSearchRq}}
    {\extend{\ldapBaseVar}{\ldapSearchVar}} \;+\;
    \standardInteraction{\protocolMdlAlphElement{?}{\ldapModRq}}
    {\extend{\ldapBaseVar}{\standardInteraction
        {\protocolMdlAlphElement{!}{\ldapModRes}}{\inaction}}} \\
    &\;+\;
    \standardInteraction{\protocolMdlAlphElement{?}{\ldapAddRq}}
    {\extend{\ldapBaseVar}{\standardInteraction
        {\protocolMdlAlphElement{!}{\ldapAddRes}}{\inaction}}} \;+\;
    \standardInteraction{\protocolMdlAlphElement{?}{\ldapDelRq}}
    {\extend{\ldapBaseVar}{\standardInteraction
        {\protocolMdlAlphElement{!}{\ldapDelRes}}{\inaction}}} \\
    \mathbf{and} & \\
    \ldapSearchVar = & \; \standardInteraction
    {\protocolMdlAlphElement{!}{\ldapSearchEntry}} {\ldapSearchVar}
    \;+\; \standardInteraction
    {\protocolMdlAlphElement{!}{\ldapSearchDone}}
    {\inaction} \\
    \mathbf{in} & \; \ldapBaseVar &
  \end{align*}
  }
\vspace{-0.75cm}
\caption{LDAP Directory Service Protocol}
\label{fig:ldap-prot}
\vspace{-0.5cm}
\end{figure}


\subsubsection{Behaviour}


The main responsibility of software services is processing requests. Clients
send requests to services which, in turn, process them and return a response
(or a sequence of responses) conveying the result. The purpose of the
behaviour meta-model is to facilitate modelling of service behaviour.


Our model allows different ways of modelling service behaviour. The specific
approach chosen depends on the level of \emph{fidelity} required of an
emulated service.
%
%
%
%
Flexibility in the behaviour meta-model is crucial so that emulated services
behave at the right level of accuracy required for the testing scenarios,
subject to the resources and service data available. We achieve this by
allowing modellers to define request handlers and bind these to the reception
of specific message shapes through a dispatch dictionary.


The dispatch dictionary contains mappings between message shapes and request
handlers. Upon receiving a request with a certain shape, the corresponding
handler from the dispatch dictionary is invoked, passing through the request
as well as the current state of the service data store. The handler returns a
sequence of messages which embody the result of the request and (optionally)
the updated state of the service data store.
Service modellers can define generic request handlers that handle requests of
many different shapes. By binding generic handlers to these different message
shapes in the dispatch dictionary, generic handlers can be reused thereby
reducing modelling effort.


The request handlers defined by service modellers need to satisfy the
function signature specified by the {\handleRqN} operation given in
Definition~\ref{def:handle-rq}. An emulated service, upon receiving a
request, 
invokes the corresponding {\handleRqN} operation, passing along the
request message for processing, paired with the current state of the
service data store. Usually, the request processing results in a
(possibly empty) sequence of responses. If the operation fails,
however, an error can be returned (omitted here for simplicity.)

\begin{defn}[Request Handlers]
  \begin{equation*}
    \handleRqN : \msgs \times \dataStores 
    \rightarrow 
    \seqOf{\msgs} \times (\emptyset \cup \dataStores)
  \end{equation*}
  where {\dataStores} denotes the set of service data stores.
  \label{def:handle-rq}
\end{defn} 

\subsubsection{Data Store}
\label{subsubsec:datastore}




Software services, such as web, email, file, database, and LDAP directory
servers, are all backed by data stores. The behaviour of each of these
services depends on the contents of these stores and the way in which that
content changes over time.
%
%
The structure and type of values within these data stores depends on
the service: Web servers providing static web pages hold structured
text of various kinds, relational databases hold tables of values of
different kinds, and LDAP directories organise their data into trees.


It is crucial, therefore, that the data store meta-model be flexible enough to
express these various structures and types. This is required for service
behaviour models to be able to accurately model the behaviour of real
services. On the other hand, it is not always necessary to have the exact same
data store representations of a modelled service as would be used in a real
service. Simplifications can reduce modelling effort, allowing shortcuts in
service behaviour models, trading fidelity for modelling effort. In order to
keep data store models as flexible as possible and to support the range of
testing scenarios mentioned, we leave their definition open and use
{\dataStores} to denote the set of data stores.

\subsection{Emulation Environment}
\label{sec:emul-env}



\begin{figure}[tb]
  \centering
  \includegraphics[width=0.7\columnwidth]{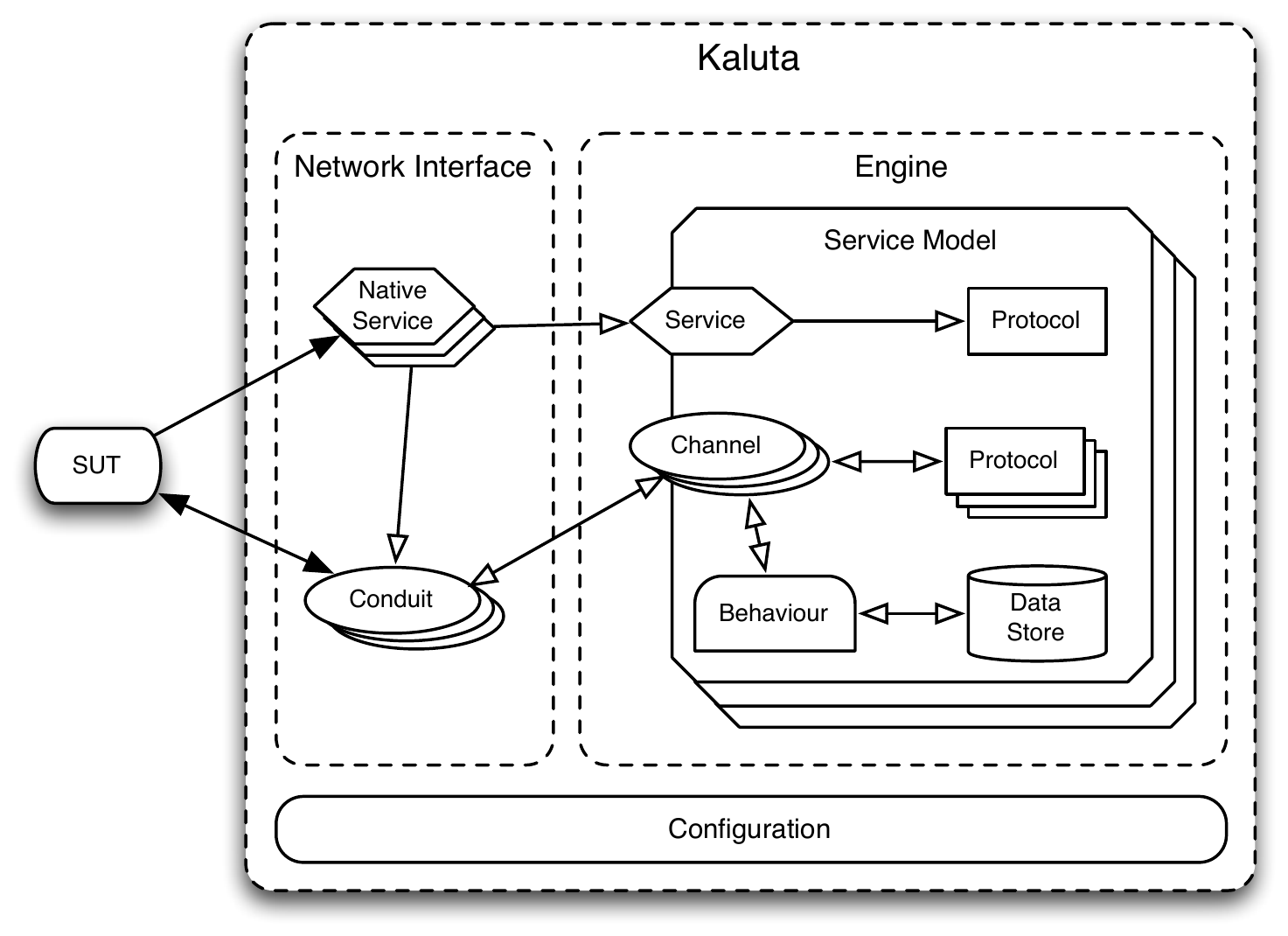}
  \caption{{\Kaluta} Architecture}
  \label{fig:kaluta-arch}
  \vspace{-0.5cm}
\end{figure}

We have constructed, {\Kaluta}, a service emulation environment, that is able
to execute service models to present the appearance and behaviour of an
enterprise software environment. {\Kaluta}'s architecture (given in
Fig.~\ref{fig:kaluta-arch}) consists of three modules: (i) a network
interface, handling communication with SUTs, (ii) an engine, executing service
models, and (iii) a configuration module to configure the network interface
and engine, respectively.

\subsubsection{Network Interface}

The network interface allows communication with SUTs in a manner which is
\emph{native} to those SUTs, that is, messages are encoded on-the-wire
according to the formatting requirements of the real service being
emulated. The network interface acts as a bidirectional translator between the
native messages transmitted over the communication infrastructure and the
internal format of emulated services.

There are two key components of the network interface: (i)
\emph{native services} which allow SUTs to establish native
communication channels with the emulator, and (ii) \emph{conduits}
which associate network channels with engine channels as well as
translating between the native message encodings required by the network
(native) channels and the messages understood by the engine. Native
services are bound to distinct IP address/port number pairs and
listen for new connection requests. Upon receiving a connection
request, the native service notifies the corresponding engine service,
forwarding the relevant details. It also constructs a conduit to
handle subsequent message exchanges. Conduits are responsible for the
exchange and transmission of messages on native channels and engine
channels. Upon receiving a native message from an SUT, a conduit
\emph{decodes} it into the message structure understood by the engine
and places it on the corresponding engine channel for
processing. Similarly, when a message sequence response is received
from an engine channel, the conduit \emph{encodes} it in the native format
and transmits it to the SUT via the corresponding native channel.

\subsubsection{Engine}

The role of the engine module is to concurrently execute multiple
service models.  We implemented the engine using the Haskell programming
language.
%
%
A service model has zero or more \emph{channels} for
communicating messages between emulated services
and external systems. Each service and channel is associated with a
corresponding \emph{protocol}. Channel protocols are maintained over
the course of an emulation by the engine to reflect valid message
receptions and transmissions. The behaviour and data store elements of
the engine's service models correspond to the behaviour and data store
layers of the meta-model. These are used by the engine to process
valid requests.

%


Fig.~\ref{alg:rq-proc} presents the algorithm used in the engine to
process message requests provided by the network interface. The
decoded message is represented by a value sequence $vs$ and
message shape $\aMsgShape$. The data store and dispatch dictionary of
the service are denoted by $d$ and $dd$, respectively. The channel's current
protocol state is denoted by $p$, while the constraint
$\exists (\aMsgShape, h) \in dd$ ensures there is a handler in
the dispatch dictionary to process messages of the request's
shape.

The first step of request processing is to retrieve the dispatch handler,
by invoking $\mathsf{lookup}$ on the dispatch dictionary and passing
in the request's message shape (Step 1). The request handler $h$ is then
invoked, passing in the request $(vs,\aMsgShape)$ and the service's data store
$d$, with the result stored in the pair $(rs,d')$ where $rs$ denotes the
response sequence and $d'$ the possibly updated state of the data store (Step
2). The message shapes of the response sequence are then used to retrieve the
corresponding continuation in the protocol model by iteratively invoking
$\mathsf{next}$ (Step 3). If the data store has been modified (i.e. $d'
\neq \emptyset$), then the service data store is updated (Step 4). Finally, the
response sequence $rs$ is returned to the network interface for encoding and
transmission (Step 5).

\begin{figure}[tb]
  \begin{algorithmic}
    \Require $(vs, \aMsgShape) \in \msgs$, $d \in \dataStores$, 
    $p \in \protocols$,\\
    \hspace{12mm}$dd \in \msgShapes \times \handleRqN$, \\
    \hspace{12mm}$\exists (\aMsgShape, h) \in dd$\\
    \vspace{-0.2cm}
    \State $h \leftarrow \mathsf{lookup}(\aMsgShape, dd)$
    \Comment{Retrieve handler (1)}

    \State $(rs, d') \leftarrow h((vs, \aMsgShape), d)$
    \Comment{Invoke handler (2)}
    
    \ForAll{$\aMsgShape'$ such that $(vs', \aMsgShape') \in rs$} 
    \State $p \leftarrow \mathsf{next}(\outSymb \aMsgShape', p)$ 
    \Comment{Update protocol (3)}
    \EndFor

    \If{$d' \neq \emptyset$}
    \State $d \leftarrow d'$
    \Comment{Update the data store (4)}
    \EndIf

    \Return $rs$ 
    \Comment{Return response sequence (5)}
  \end{algorithmic}
  \caption{Request Processing}
  \label{alg:rq-proc}
  \vspace{-0.4cm}
\end{figure}





\section{Evaluation}
\label{sec:evaluation}

{\Kaluta} was evaluated with respect to three research questions:
(\emph{RQ1}) can {\Kaluta} emulate \numprint{10 000} endpoints on a
single physical host and how does scale affect resource consumption
(CPU computation and memory usage)? (\emph{RQ2}) how does {\Kaluta}'s
resource consumption compare to one of the most common alternative
approaches -- VMs? (\emph{RQ3}) what unique benefits can {\Kaluta}
bring to the testing enterprise software systems?


\subsection{Scalability of {Kaluta} (RQ1)}
\label{ss:kal-scalability}

\subsubsection{RQ1 -- Experimental Setup}
\label{ss:kal-method}

A workload script was written to invoke a series of operations on LDAP
directories, typical of the types of operations an identity management system
performs in an enterprise environment. The sequence of operations were as
follows: (i) open a network connection, (ii) bind to the LDAP directory, (iii)
retrieve the whole directory through search, (iv) add a new user, (v) search a
particular sub-tree of the directory, (vi) modify a user's password, (vii)
search for a specific entry of the directory (verifying the preceding password
modification), (viii) delete a user, and finally (ix) unbind. The workload
script was executed with 32 concurrent user threads. Within each user thread,
requests were sent synchronously.




{\Kaluta} was installed on a Dell PE2950 server, with dual quad-core Intel
Xeon E5440 2.83GHz CPUs and 24GB of RAM. The workload script was executed on
another machine which had a dual core Intel Pentium 4 CPU and 2GB of RAM.
Both machines ran the Ubuntu 11.04 64-bit as their operating systems. The two
machines were connected via a 1 Gigabit/s Ethernet connection.

Each directory server model was initialised with a data store containing 100
entries. The number of LDAP servers emulated was varied between 1 and
\numprint{10 000}. Each emulated endpoint was given a separate IP address
using the Linux \texttt{ifconfig} utility. CA Application Performance
Management (APM) version 9.1 monitored {\Kaluta}'s CPU consumption, memory
usage, and the time taken to process workloads. Each experimental
configuration was run at least 6 times~\cite{Box1987}.

\subsubsection{RQ1 -- Results}
\label{ss:kalresults}

\begin{figure}[tb]
  \centering
  \caption{RQ1 Results}
  \begin{tabular}{c}
    \multicolumn{1}{c}{\includegraphics[width=0.72\columnwidth]
      {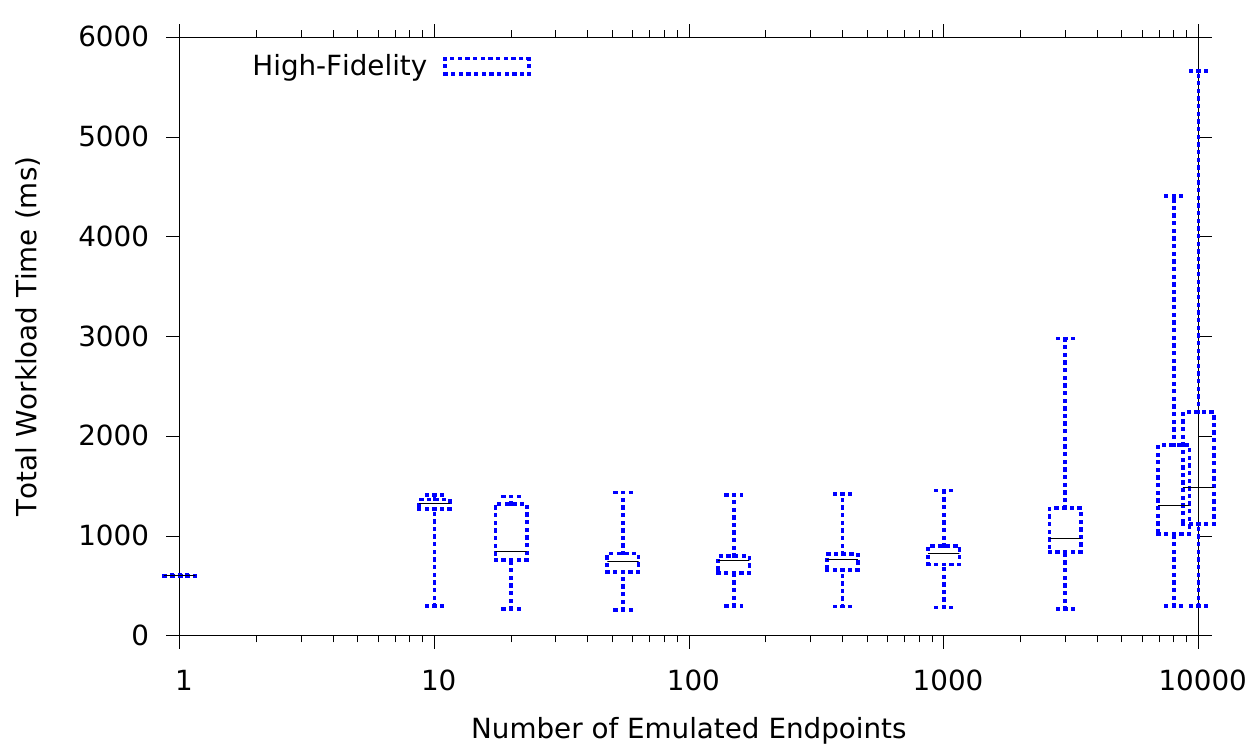}}\\
    \multicolumn{1}{c}{\scriptsize (a) Workload Processing Times}\vspace{2mm}\\
    \includegraphics[width=0.72\columnwidth]{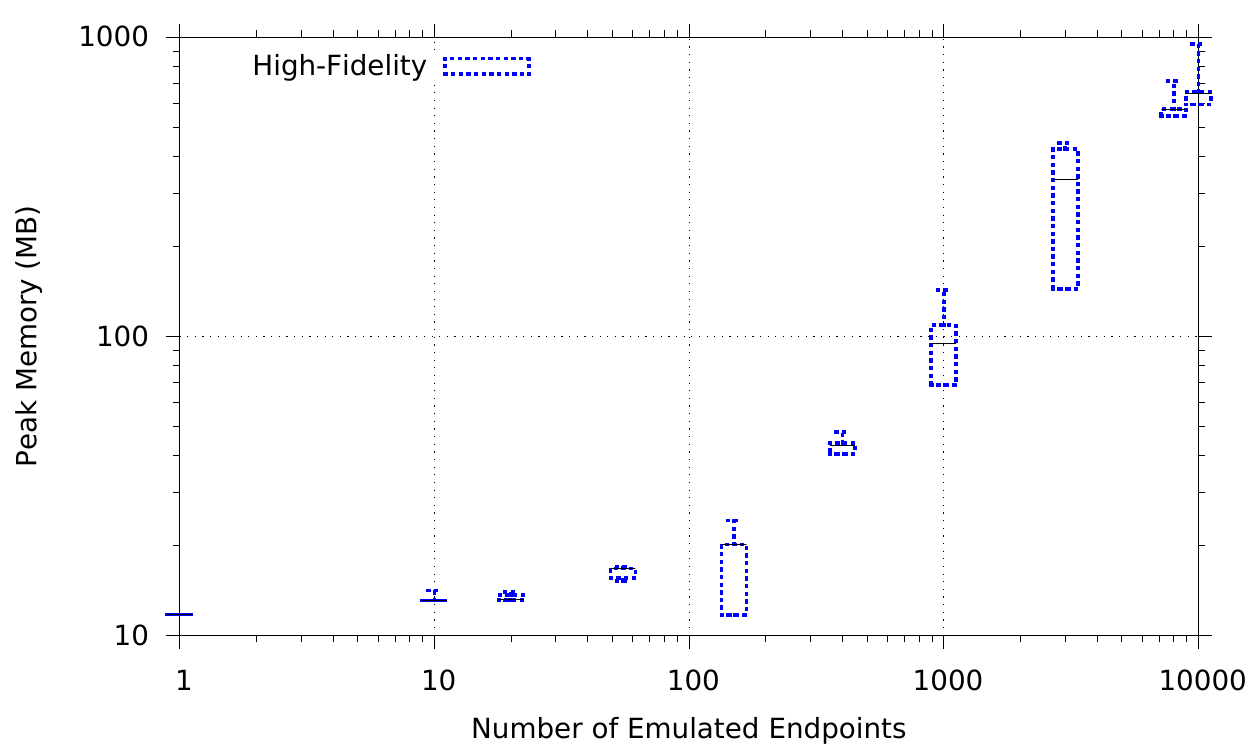}\\
    {\scriptsize (b) {\Kaluta} Peak Memory - Engine Module}\\
    \includegraphics[width=0.72\columnwidth]{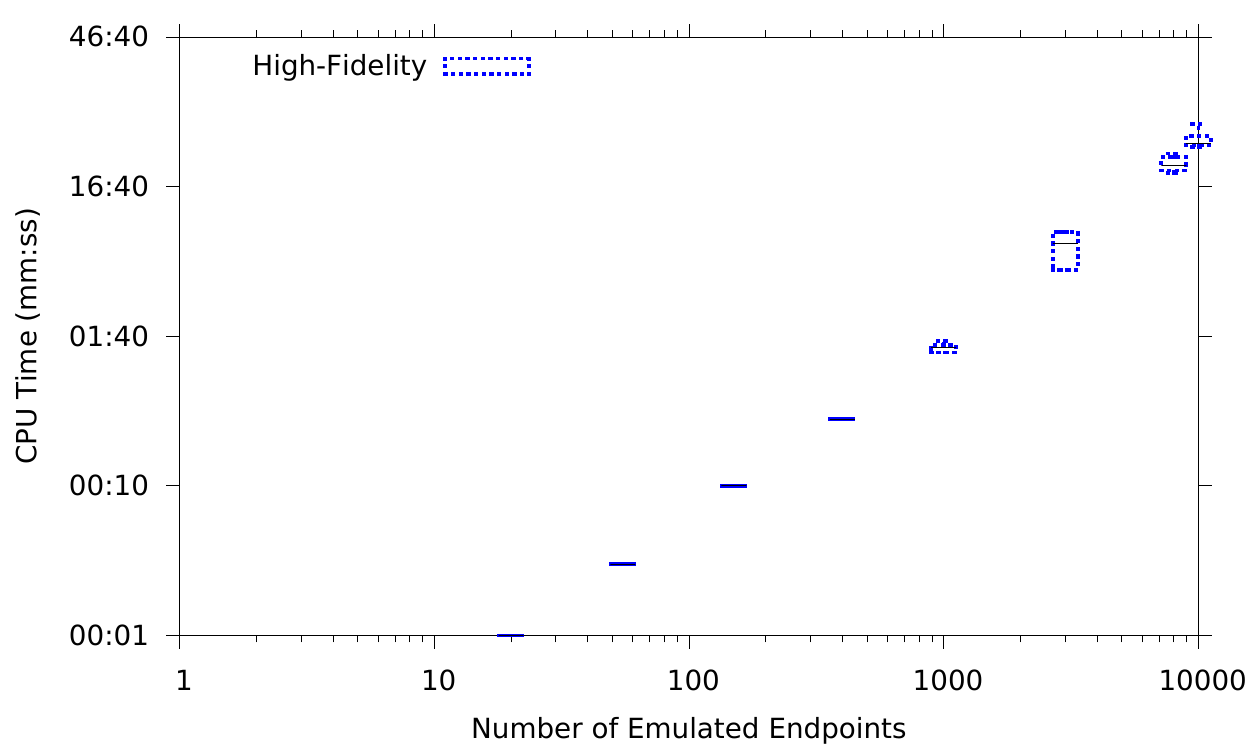}\\
    {\scriptsize (c) {\Kaluta} Total CPU Usage}
  \end{tabular}
  \label{fig:rq1}
  \vspace{-0.4cm}
\end{figure}

{\Kaluta} successfully emulated \numprint{10 000} LDAP servers on a single
physical host. The elapsed time it took {\Kaluta} to process a workload for
increasing numbers of endpoints is shown in Fig.~\ref{fig:rq1}(a).  For each
emulated endpoint, about 125 LDAP messages were exchanged between the SUT and
the endpoint. For \numprint{1 000} endpoints or less, the median workload
processing time was 748 milliseconds. For over \numprint{2 000} endpoints,
workload processing times increased and there was also greater variability in
the workload processing times. The median workload processing time per
endpoint for \numprint{10 000} emulated endpoints was about 50\% slower
compared to the median processing times for emulations of \numprint{1 000}
endpoints or less. Despite the performance degradation, {\Kaluta}'s response
times, even for \numprint{10 000} endpoints, was fast enough to not be the
bottleneck in the testing of {\IM} ({\IMShort}).
Section~\ref{ss:jcs-experiment} describes how {\Kaluta} was able to generate
responses faster than {\IMShort} was able to generate requests.

Figures ~\ref{fig:rq1}(b) and \ref{fig:rq1}(c) show the memory consumption and
CPU usage of {\Kaluta}, respectively. The peak memory consumption of {\Kaluta}
for \numprint{10 000} LDAP endpoints was about 650MB. {\Kaluta}'s engine has a
peak memory usage and total computation time that increased linearly with the
number of emulated endpoints: processing the workloads for \numprint{10 000}
endpoints consumed a total of 32 minutes of CPU time (spread across up to 8
cores of the host machine.)




\subsection{Comparison to VMs (RQ2)}
\label{ss:vmcomp}

\subsubsection{RQ2 -- Experimental Setup}
The VM experiments were conducted using the same workload script, hardware and
environment as for RQ1. VMware Player version 4.0 was used as the hypervisor. A
VM was created for each instance of an LDAP server endpoint. The VMs were run
on the same host machine as {\Kaluta}. For each VM we did a minimum install
of Ubuntu 11.04 64-bit and installed OpenLDAP Server. Each VM was allocated
128MB of main memory (the minimum needed for the VM to boot) and was given a
10GB virtual hard disk, which occupied about 1.7GB of physical disk space.

The recommended upper limit of virtual CPUs per physical core is between 8 and
10 \cite{Sanchez:11}. Since the host machine had 8 physical cores, in order to
ensure a high performance from the VMs, we stayed within the recommended limit
and ran 60 single virtual CPU VMs. The workload was then applied to the 60
OpenLDAP Server applications running on separate VMs. We used APM to monitor
the resource usage of the VMs.

\subsubsection{RQ2 -- Results}


The comparison of the resource consumption of 60 VM endpoints to that of 60
{\Kaluta} emulated endpoints is given in Table~\ref{tab:vm-comparison}. The
resource usage of {\Kaluta} is order of magnitudes less than that of the VM
endpoints.  With respect to peak memory usage, {\Kaluta} uses about 66 times
less.  In terms of CPU consumption, {\Kaluta} uses about 90 times less.
Finally, with respect to the hard disk space consumed, 60 VMs occupied over
100GB whereas the disk space taken by the {\Kaluta} models and configuration
files was negligible.

\begin{table}[tb]
  \caption{Resource usage: VMs versus {\Kaluta}.}
  \label{tab:vm-comparison}
  \centering
  \begin{tabular}{|l|r|r|}
    \hline
	& {VMs} & {\Kaluta} \\
    \hline
    Peak memory usage     & \numprint{11 006} MB & 162 MB \\
    Total CPU consumption & 826.69 ms & 9.29 ms \\
    Hard disk space       & 102 GB     & 65 KB \\
    \hline
  \end{tabular}
  \vspace{-0.4cm}
\end{table}




\subsection{Scalability Testing of {\IMShort} (RQ3)}
\label{ss:jcs-experiment}

\subsubsection{RQ3 -- Experimental Setup}

We used {\Kaluta} to evaluate the scalability of {\IM} (cf.
Section~\ref{sec:ind-scen}). The first requirement for {\Kaluta} is that its
models are accurate enough to `fool' {\IMShort} that it is interacting
with real endpoints, indicating that its responses need to be consistent with
the those {\IMShort} expects from real endpoints. To validate this, we used
{\IMShort}'s user interface to acquire a {\Kaluta} emulated LDAP endpoint,
explore it, add and modify some users. {\IMShort} was able to perform these
operations without errors, indicating that the emulated endpoints behaviour
was consistent with {\IMShort}'s expectations.

We then created an experiment to measure {\IMShort}'s scalability when
managing up to \numprint{10 000} endpoints. The CA {\em IAM Connector Server}
(CS) is the component of {\IMShort} which communicates with the endpoints and,
therefore, requires the greatest scalability. We installed a development
version of the CS on a separate machine to {\Kaluta}. The CS machine had a
quad-core Intel Xeon X5355 CPU, 12GB of RAM, and ran Windows Server 2008 R2
64-bit as operating system. The only change made to the CS configuration from
the installation defaults was to increase the maximum heap size to 5GB. The CS
machine and the {\Kaluta} machine were connected via a 1 Gigabit/s Ethernet
connection. {\Kaluta} was configured to emulate \numprint{10 000}
LDAP endpoints.

A JMeter script was written to automate the CS to invoke the same set of
identity management operations on each endpoint as described in
Section~\ref{ss:kal-method}. The script was run with different numbers of
concurrent user threads, varied between 1 and 100. For each number of threads,
the experiment was run six times (and the results averaged). APM was again
used to monitor the memory and CPU usage.


\subsubsection{RQ3 -- Results}

The test ran successfully and demonstrated that a single instance of the CS
could manage in excess of \numprint{10 000} endpoints, a scale previous
unachieved by the {\IMShort} developers during testing. The average completion
times for varying numbers of JMeter user threads is given in
Table~\ref{tab:jthreads}. The single threaded experiment took 4.5 hours to
complete. Using 20 user threads reduced the execution time to about 82
minutes. Adding more than 20 user threads did not further reduce the execution
time of the experiment. These were acceptable completion times for acquiring
and updating \numprint{10 000} endpoints, especially considering the hardware
configuration that was used for running the experiments.

\begin{table}[tb]
  \centering
  \caption{Execution times for \numprint{10 000} endpoints.}
  \begin{tabular}{rcccccc}
    \toprule
    JMeter Threads&1&2&5&10&20&100\\
    Total Time (min)&272.8&141.3&95.8&85.6&81.9&82.3\\
    \bottomrule
  \end{tabular}
  \label{tab:jthreads}
  \vspace{-0.4cm}
\end{table}



\begin{figure}[bt]
  \caption{Connector Server Resource Consumption}
  \begin{tabular}{cc}
    \includegraphics[width=0.5\columnwidth]{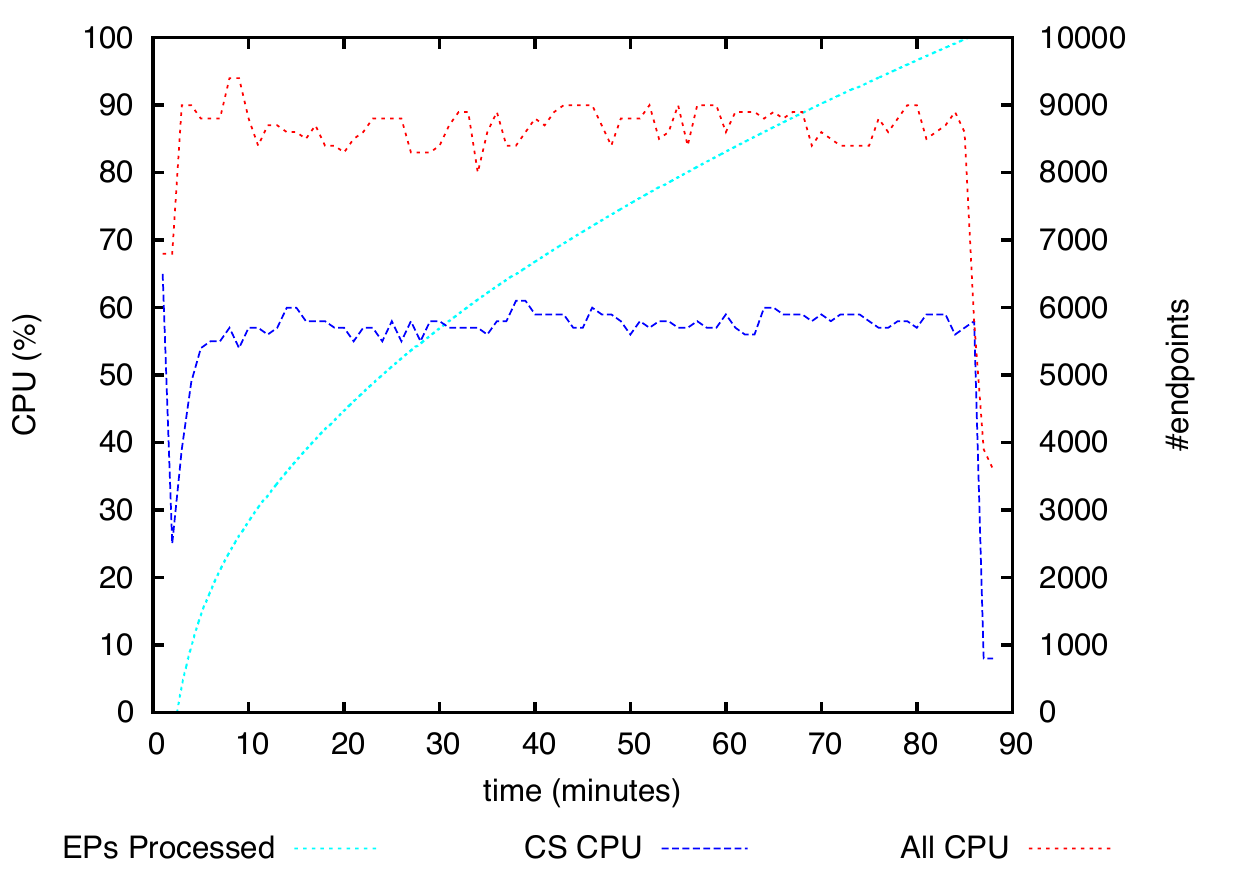}&
    \includegraphics[width=0.5\columnwidth]{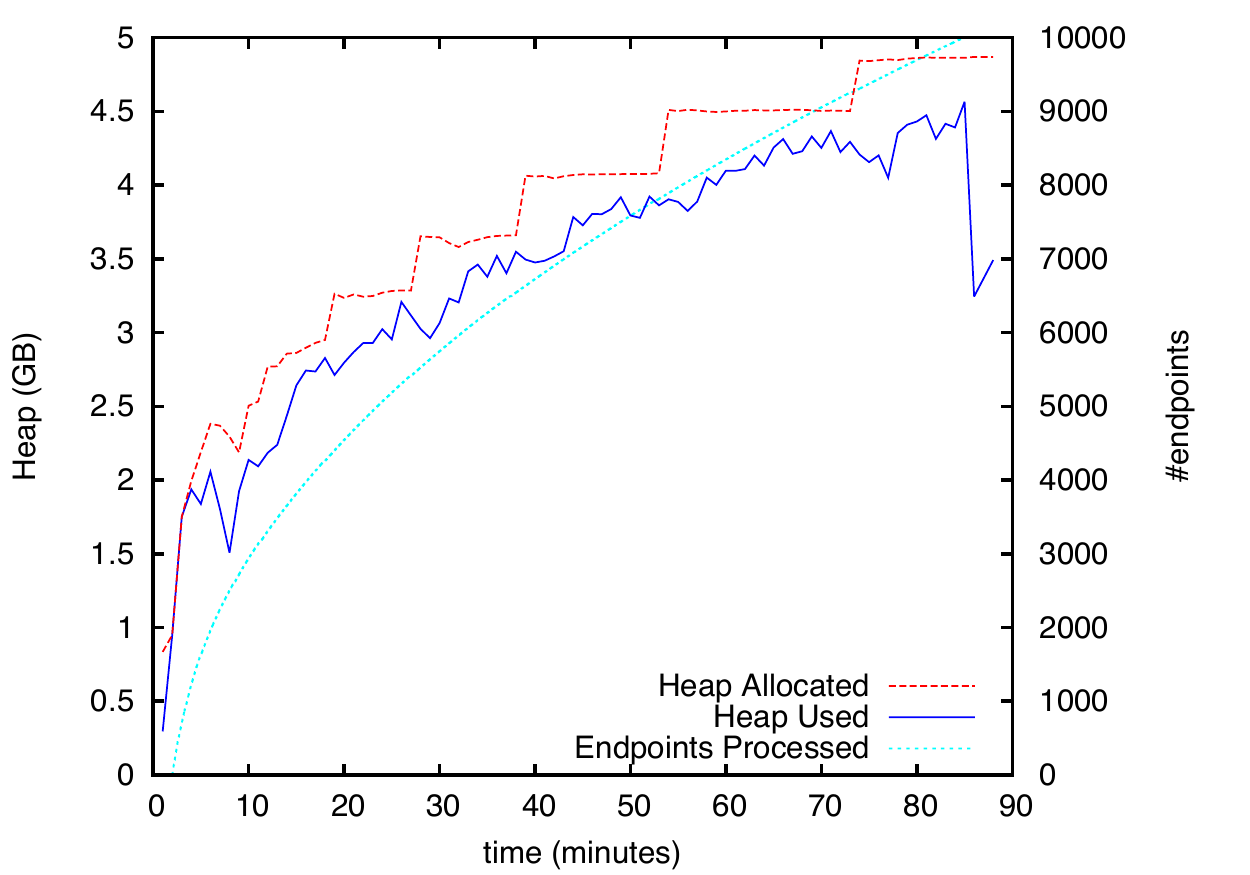}\\
    {\scriptsize (a) CPU}&
    {\scriptsize (b) Heap}
  \end{tabular}
  \label{fig:rq3}
  \vspace{-0.4cm}
\end{figure}

Figures \ref{fig:rq3}(a) and \ref{fig:rq3}(b) show the CPU utilisation and
heap usage, respectively, of the CS throughout the course of a sample run with
20 user threads. The CS used 50-60\% CPU over the course of the experiment.
When all processes were included, total CPU usage sometimes reached up to
90\%. The heap size of the CS steadily grew as more endpoints came under
management. When the full \numprint{10 000} endpoints were acquired, the heap
utilisation was about 4.5GB.

{\Kaluta} was able to handle all incoming requests generated by the CS, and
its memory and CPU usage were consistent with the results reported in
Section~\ref{ss:kalresults}. For each run of the experiment,
app. \numprint{270 000} LDAP messages were exchanged. {\Kaluta}'s conformance
checker confirmed that the CS sent no messages outside of the allowable
protocol sequence.

%
%

We also ran some experiments to test how the CS handles protocol
non-conformance from the endpoints. For example, we ran tests where {\Kaluta}
randomly delayed its responses by up to 30 seconds, or did not respond at all.
The CS handled both the delayed responses and non-responses correctly.


\subsection{Unique Insights using Kaluta}
\label{ss:kal-benefits}




{\Kaluta} gave multiple insights into the characteristics of the {\IM} system,
and we were able to confirm that {\IMShort} scales to manage at least
\numprint{10 000} endpoints using only a single instance of the Connector
Server (CS).

{\Kaluta} allowed us to observe the resource consumption of the CS component
of {\IMShort} while operating at large scale. This profiling information was
passed on to the software developers, giving them information which could not
be easily obtained through other forms of testing. The memory consumption at
large scales was higher than expected, and further investigations revealed
issues with connection caching and logging, respectively. The software
developers used this information to improve the design of the CS, increasing
performance by a factor of 100, and reducing memory consumption by 80\% for
the \numprint{10 000} endpoint scale. Only by running tests at large scales
could these issue be found and resolved. We were further able to demonstrate
protocol conformance of CS as none of the messages exchanged violated the
expectations of the underlying protocol.


By collecting information about resource consumption and performance at large
scales, we were able to provide guidelines to IT implementers with respect to
the system resources which will be required for a deployment of a given size.
%
%
Finally, our testing revealed that for large-scale deployments, the operating
environment itself needs to be validated as well. For example, when we
deployed our workload generation script on Ubuntu Linux, the size of the ARP
(Address Resolution Protocol) cache table needed to be increased, in order to
achieve a timely completion of the test. This was due to connecting to a
greater number of different IP addresses than a default Linux configuration
allows.

\section{Related Work}
\label{sec:related-work}


There is work describing conceptual~\cite{Colombo2005,Quartel2007} and
formal~\cite{Broy2007} service models. Colombo \etal\cite{Colombo2005} model
services in terms of the core agents, actors and their relationships to one
another. Our context, expressed using the terminology of their model, consists
of two primary agents, the SUT and the testbed, playing the roles service
consumer and service provider, respectively. {\Kaluta} provides concrete
simple services to the SUT which are lower fidelity than real services but
suitable for many testing scenarios. Quartel \etal\cite{Quartel2007} present
COSMO, a conceptual service modelling framework supporting refinement. Our
service model can be interpreted as focusing on a subset of the service
aspects and abstraction level presented in COSMO. Namely a service model which
focuses on the behavioural and information aspects of services at the
choreography level of abstraction.


The role of {\Kaluta} is similar to that of {\sc{Puppet}}~\cite{Bertolino2008}
that uses a model-based approach to generate stubs for Web Services. The
functional behaviour of emulated Web Services are defined as Symbolic
Transition Systems (STSs), or by UML 2.0 state machines which are translated
into STSs for execution. The STS models on which functional behaviour is based
encompasses the temporal, logic, and state aspects of service behaviour within
a single model. Our service meta-model segregates these three aspects into
separate layers: the protocol, behaviour and data store layers, respectively.
This allows the possibility of different models to capture these properties to
be mixed in at a later stage if beneficial. A protocol model based on
Petri-nets 
or linear time logic may, for instance, be incorporated into later {\Kaluta}
versions. Another difference between {\sc{Puppet}} and our own work is our
focus on scalability. Although {\sc{Puppet}} incorporates qualities through
service level agreements into the testbed, it does not set out to represent
environments containing thousands of concurrent services.





\section{Conclusions and Future Work}
\label{sec:conclusion}

The scale of some large distributed environments makes it quite difficult to
construct testbeds representative of production conditions.
%
%
Service emulation is an approach we propose to constructing large scale
testbeds. We have presented (i) a layered service meta-modelling framework
facilitating service modelling up to a flexible level of fidelity,
accommodating the needs of different testing scenarios; (ii) {\Kaluta}, an
emulation environment supporting scalable service model execution and
interaction with external SUTs; and (iii) an empirical evaluation
investigating the scalability and resource consumption of {\Kaluta}, comparing
it with VM approaches, and investigating its effectiveness in industry testing
scenarios. We find that {\Kaluta} is substantially more scalable than VM
approaches, capable of emulating \numprint{10 000} LDAP directory servers
using a single physical host. Furthermore, {\Kaluta} was used to gain unique
insights into the run-time properties of a real enterprise software system, CA
{\IM}, when operating at large scales. This insight allowed the design of
{\IMShort} to be improved, boosting its performance in large scale conditions
by a factor of 100.

Future work includes automating aspects of service model synthesis to reduce
human modelling effort. We will also investigate how to explicitly specify
time delays in endpoint models to better cater for services with significant
computational complexity. Finally, we plan to connect a network emulator to
{\Kaluta} to investigate effects of various network settings and topologies on
SUTs, respectively.


\subsection*{Acknowledgments}

We would like to thank the Australian Research Council for their support
of this work under grant LP100100622.


\bibliographystyle{abbrv}
\bibliography{./caise-2013}

\end{document}